\documentclass{article}
\usepackage{ismir,amsmath,cite,url}
\usepackage{graphicx}
\usepackage{color}
\usepackage[utf8]{inputenc} 
\usepackage[T1]{fontenc}    
\usepackage{hyperref}       
\usepackage{url}            
\usepackage{booktabs}       
\usepackage{amsfonts}       
\usepackage{nicefrac}       
\usepackage{microtype}      
\usepackage{xcolor}

\usepackage{amsmath}
\usepackage{enumerate}
\usepackage{array} 
\usepackage{algorithm}
\usepackage[noend]{algpseudocode}
\usepackage[normalem]{ulem}
\usepackage{caption}
\usepackage{subcaption}
\usepackage[toc,page]{appendix}

\usepackage{changepage}

\newcommand{\rulesep}{\unskip\ \vrule\ }

\title{Ensemble-based cover song detection}

\threeauthors
  {Marc Sarfati} {Spotify \\ {\tt marcs@spotify.com}}
  {Anthony Hu} {Spotify \\ {\tt anthonyh@spotify.com}}
  {Jonathan Donier} {Spotify \\ {\tt jdonier@spotify.com}}

\begin{document}

\maketitle
\begin{abstract}

Audio-based cover song detection has received much attention in the MIR community in the recent years. To date, the most popular formulation of the problem has been to compare the audio signals of two tracks and to make a binary decision based on this information only. However, leveraging additional signals might be key if one wants to solve the problem at an industrial scale. In this paper, we introduce an ensemble-based method that approaches the problem from a many-to-many perspective.  Instead of considering pairs of tracks in isolation, we consider larger sets of potential versions for a given composition, and create and exploit the graph of relationships between these tracks. We show that this can result in a significant improvement in performance, in particular when the number of existing versions of a given composition is large.

\end{abstract}


\section{Introduction}

With the rise of online streaming services, it is becoming easier for artists to share their music with the rest of the world. With catalogs that can reach up to tens of millions of tracks, one of the rising challenges faced by music streaming companies is to assimilate ever-better knowledge of their content -- a key requirement for enhancing user and artist experience. From a musical perspective, one highly interesting aspect is the detection of composition similarities between tracks, often known as the \textit{cover song} detection problem. This is, however, a very challenging problem from a content analysis point of view, as artists can make their own version of a composition by modifying any number of elements -- instruments, harmonies, melody, rhythm, structure, timbre, vocals, lyrics, among others. 

Over the years, it has become customary in the Music Information Retrieval (MIR) literature to address the cover song detection problem in what is arguably the most challenging setting. Indeed, most papers attempt to detect composition relationships between pairs of tracks based on their two audio signals only -- in other words, completely out of context and without using any metadata information. While this well-defined task makes sense from an academic perspective, it might not be the optimal approach for solving the problem at an industrial scale  \cite{correya2018large}.

The second starting point of our work is the fact, often mentioned in cognitive science, that commonly observed patterns are represented and stored in a redundant fashion in the human brain, which makes them more likely to be retrieved, recognised and identified than patterns that are observed less frequently \cite{kurzweil2013create}. If true, this would apply to our assessment of composition similarities as well. The main idea behind our work is that the corpus of existing versions of a composition can be precisely a substitute for these multiple representations. 

Following these guiding intuitions, we turn to a new use case, where we do not just have pairs but a pool of \emph{candidates} that are likely to be instances of some given musical work (according \textit{e.g.} to some first metadata analysis). We then compare these candidates not only to \emph{one} reference version (\textit{e.g.} the original track, if it exists) but also to other candidate versions. We then build a graph of all these versions to identify composition clusters. Sometimes, when hundreds or thousands of versions of a given work exist (which is quite common in the catalogue of a streaming company), this ensemble-based approach can result in substantial improvements on the cover detection task.

In Section \ref{sec:literature} we present a review of the literature on cover identification. In Section \ref{sec:1v1}, we present the 1-vs-1 cover identification algorithm that we use throughout the paper, which is heavily based on \cite{tralie2017early}. 
The main contribution of this paper lies in Section \ref{sec:clustering}, in which we present the new use case for cover identification described in the previous paragraph.
We then showcase our method with examples in Section \ref{sec:results} and discuss some challenges in Section \ref{sec:discussion}. 


\section{Related work}
\label{sec:literature}

A number of possible approaches for cover song identification have been developed in the last decade, with varying levels of performance. Reference \cite{ellis2007identifyingcover} introduced a first solution to this problem and has been used as a starting point for many subsequent studies. The main idea is to extract a list of beat-synchronous \cite{durand2017robust} chroma features from two input tracks and quantify their similarity by applying dynamic programming algorithms to a cross-similarity matrix derived from these features. This algorithm has been refined in \cite{ellis20072007} by the same authors by adding a few modifications such as tempo biasing to improve the results. 
Harmonic Pitch Class Profile (HPCP) features (chroma features) have proven very useful in cover identification \cite{ellis2007identifyingcover, serra2008chroma, ravuri2010cover, serra2008transposing} as they capture meaningful musical information for composition. 
Other features have subsequently been introduced
, such as self-similarity matrices (SSM) of Mel-Frequency Cepstral Coefficients (MFCC) \cite{tralie2015cover, tralie2017early}. To take advantage of the complementary properties of different types of features, \cite{tralie2017early} further introduced a method to combine several audio features by fusing the associated cross-similarity matrices, which resulted in a significant increase in performance compared to single-feature approaches. 

Having extracted audio features from two tracks to be compared, most methods use dynamic programming (either Dynamic Time Warping or the Smith-Waterman algorithm \cite{smith81}) to assign a score to the pair \cite{tralie2017early, tralie2015cover, ellis2007identifyingcover}. One drawback of these methods is that they are computationally expensive and cannot be run at scale. Hence other authors have developed solutions that enable cover identification at scale by mapping audio features to smaller latent spaces. For instance, \cite{bertin2012large, humphrey2013data} use Principal Component Analysis (PCA) to compute a condensed representation of audio features 
which they use to perform a large-scale similarity search (\textit{e.g.} a nearest neighbor search). In the same vein, references \cite{fang2017deep, qi2017audio} use deep neural networks to learn low-dimensional representations of chroma features.




\section{Pairwise matching}
\label{sec:1v1}

As mentioned above, our ensemble-based cover identification method consists of two steps. For a given work, we proceed to:
\textit{\textbf{(i)}} a pairwise (1-vs-1) comparison of all the tracks in a pool of potential candidates, 
\textit{\textbf{(ii)}} a clustering of these candidates based on the results of step (i). 

In this section we present the 1-vs-1 cover song identification algorithm (i) which will be used as a starting point for our ensemble-based approach, and evaluate its performance on two distinct cover datasets.

\subsection{The algorithm}
\label{subsec:1v1}

\begin{figure}[!ht]
\centering
 \includegraphics[width=0.5\textwidth]{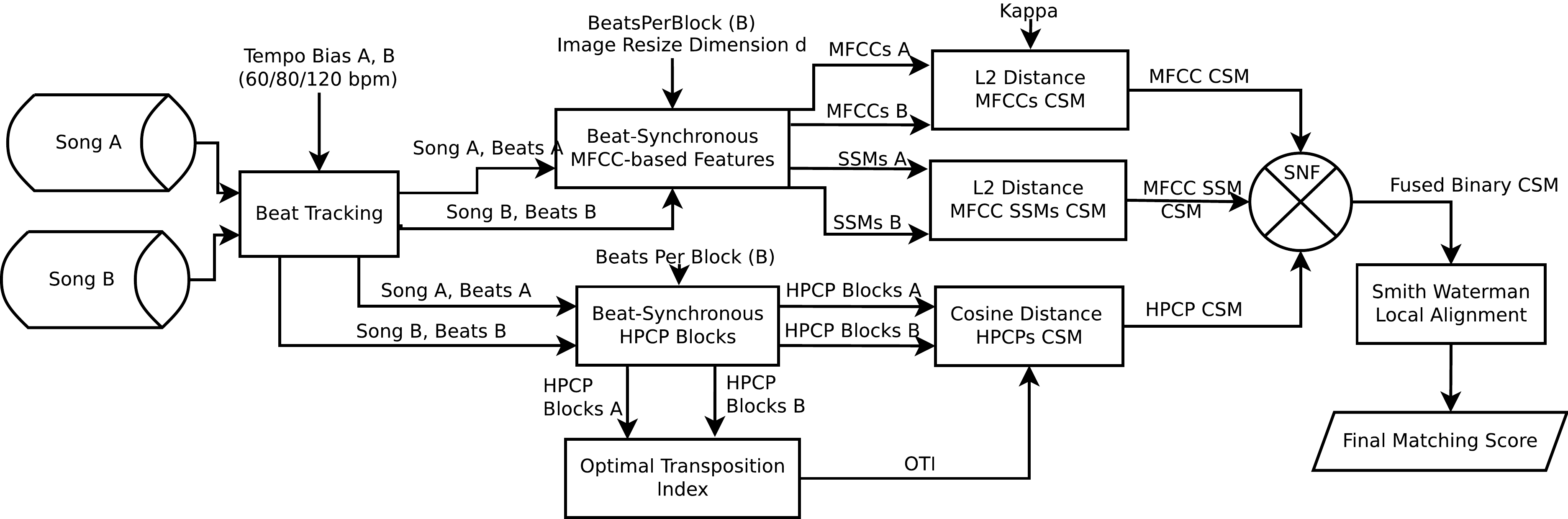}
\caption{\label{fig:pipeline} High level overview of the 1-vs-1 matching pipeline. Image reproduced from \cite{tralie2017early}.}
\end{figure}

For the purposes of this work, any 1-vs-1 similarity measure could be used for step (i), as we are mainly interested in quantifying the impact of step (ii) on the overall performance. We have chosen to rely on an implementation of the algorithm introduced in \cite{tralie2017early}, as the algorithm achieves the best results to date on the \textit{Covers80} \cite{ellis2007covers80} and \textit{MSD} (\textit{Covers1000}) datasets \cite{bertin2011million}. 
A high-level overview of the pipeline is shown on Figure \ref{fig:pipeline}. As with most algorithms presented in Section \ref{sec:literature}, it can be decomposed into two stages: first, it extracts a list of meaningful audio features from the two tracks to be compared, then it computes a similarity score based on these. The details of this method are not directly relevant to our work, so we will focus here on a quantitative assessment of its performance, to give the reader a quantitative idea of our starting point. For those interested in the details of how the algorithm works, please refer to \cite{tralie2017early}.

\subsection{Quantitative evaluation of the 1-vs-1 method}
\label{subsec:results}


We evaluate our implementation of \cite{tralie2017early} on two different datasets, and compare it with the numbers reported in the original paper as well as with a publicly available implementation of \cite{tralie2017early} by its authors.\footnote{https://github.com/ctralie/GeometricCoverSongs} To make the comparison more interpretable, we evaluate two versions of our implementation with two sets of parameters: \textit{Params1} mimics the parameters used in \cite{tralie2017early}, and should therefore produce numbers that very similar to those described in the original paper, while \textit{Params2} uses shorter 8-beats-long blocks.

We first compare the algorithms on the widely used \emph{Covers80} dataset \cite{ellis2007covers80} to enable comparison with other published methods. The dataset is composed of 160 tracks that are divided into two sets (A and B) of 80 tracks each, with every track in set A matching one (and only one) track in set B. For each of the 160 tracks, we compute its score with all the other 159 tracks and report the rank of its true match. Table \ref{table:benchmark} reports the Mean Rank (MR) of the true match (1 is best), the Mean Reciprocal Rank (MRR) \cite{craswell2009mean}, as well as the Recall@1 (R@1) and Recall@10 (R@10). We also compute the so-called \emph{Covers80 scores} by querying each track in set A against all the tracks in set B and reporting the number of matches found with rank 1.\footnote{Each track from set A is now queried against the 80 tracks from set B, instead of all other 159 tracks.} Overall, our results are close to the ones reported in \cite{tralie2017early} -- even though we could not quite reach the numbers given in their paper.

\begin{table}[ht!]
\hspace{-4ex}
\footnotesize
\begin{center}
\setlength{\tabcolsep}{0.78ex}
\begin{tabular}{| c | c | c | c | c | c || c | c |}
  \hline	
  & \multicolumn{5}{c ||}{\textbf{Covers80}} & \multicolumn{2}{c |}{\begin{tabular}[x]{@{}c@{}}\textbf{Internal}\\ \textbf{Dataset} \end{tabular} }\\
  \cline{2-8}
  & MR & MRR & \begin{tabular}[x]{@{}c@{}}R@1\\(/159) \end{tabular} & \begin{tabular}[x]{@{}c@{}}R@10\\(/159) \end{tabular} & \begin{tabular}[x]{@{}c@{}}\emph{Covers80}\\score \end{tabular} & Recall & \begin{tabular}[x]{@{}c@{}}Recall\\(no Jazz) \end{tabular} \\
  \hline
  \cite{tralie2017early} paper & \textbf{7.8} & \textbf{0.85} & \textbf{131} & 143 & \textbf{68/80} & - & - \\
  \cite{tralie2017early} code & 8.6 & 0.77 & 114 & \textbf{145} & 62/80 & 73.2\% & 81.8\%  \\
  Params1 & 10.5 & 0.81 & 125 & 136 & 64/80 &  79.2\% & 87.8\% \\
  Params2 & 13.2 & 0.75 & 116 & 128 & 60/80 & \textbf{85.8\%} & \textbf{95.1\%} \\
   \hline  
  \end{tabular}
 \end{center}
 \caption{\label{table:benchmark} Comparison of our implementations (\textit{Params1} and \textit{Params2}) against the implementations of \cite{tralie2017early}, on the \textit{Covers80} dataset and on our internal dataset. 
 The recall rates for the internal dataset correspond to a false positive rate of 0.5\%. For each column, the best performance is printed in bold.
 }
\end{table}

To complement this baseline, we have created an internal dataset of 452 pairs of covers grouped into several categories, obtained by metadata filtering based on the keywords \textit{Acoustic Cover, Instrumental Cover, Karaoke, Live, Remix, Tribute} as well as some \textit{Classical} and \textit{Jazz} covers. Such granularity allows us to compare the performance of our algorithm across genres and cover types, providing a new perspective on the problem, as shown in Table \ref{table:recall}. We have tested the two versions of our algorithm on the 452 positive pairs and 10,000 negative pairs selected uniformly at random. We selected the classification threshold to ensure a very low false positive rate below $0.5\%$. Results are presented in Table \ref{table:benchmark}. Our algorithm reaches $86.3\%$ recall, versus $73.2\%$ for the publicly available implementation of \cite{tralie2017early}\footnote{As the computational time is much higher for this algorithm, we only computed the false positive rate using 500 negative pairs.}. Note that jazz is the most challenging genre to detect, as jazz covers include a lot of improvisation that can be structurally different from their parent track (see Table \ref{table:recall}). 
If we remove jazz covers from the dataset, the recall increases to $95.1\%$ with the \emph{Params2} implementation.



\begin{table}[h!]
\begin{adjustwidth}{-0.5cm}{}
\footnotesize
\begin{center}
\setlength{\tabcolsep}{0.7ex}
\begin{tabular}{| c | c c c c c c c c |}
  \hline	
  \textbf{Type} & Acoustic & Instr. & Karaoke & Live & Remix & Tribute & Classical & Jazz\\
  \hline	
  \textbf{\# of pairs} & 57 & 63 & 46 & 57 & 31 & 53 & 77 & 68 \\
  \hline	
  \textbf{Recall} & 94\% & 84 \% & 97\% & 100\% & 93 \% & 96 \% & 100 \% & 35 \% \\
  \hline	
  \end{tabular}
 \end{center}
 \end{adjustwidth}
 \caption{\label{table:recall} Recall rates for each genre in our internal dataset, with a $<0.5\%$ false positive rate.}
\end{table}

In view of these results, we will use our own implementation with \textit{Params2} throughout the rest of this paper, as it is faster and performs best on our internal dataset, which is larger and more diverse than \textit{Cover80}.

\subsection{Distributions of scores}
\label{subsec:threshold}

\begin{figure}[!ht]
\centering
 \includegraphics[width=.45\textwidth]{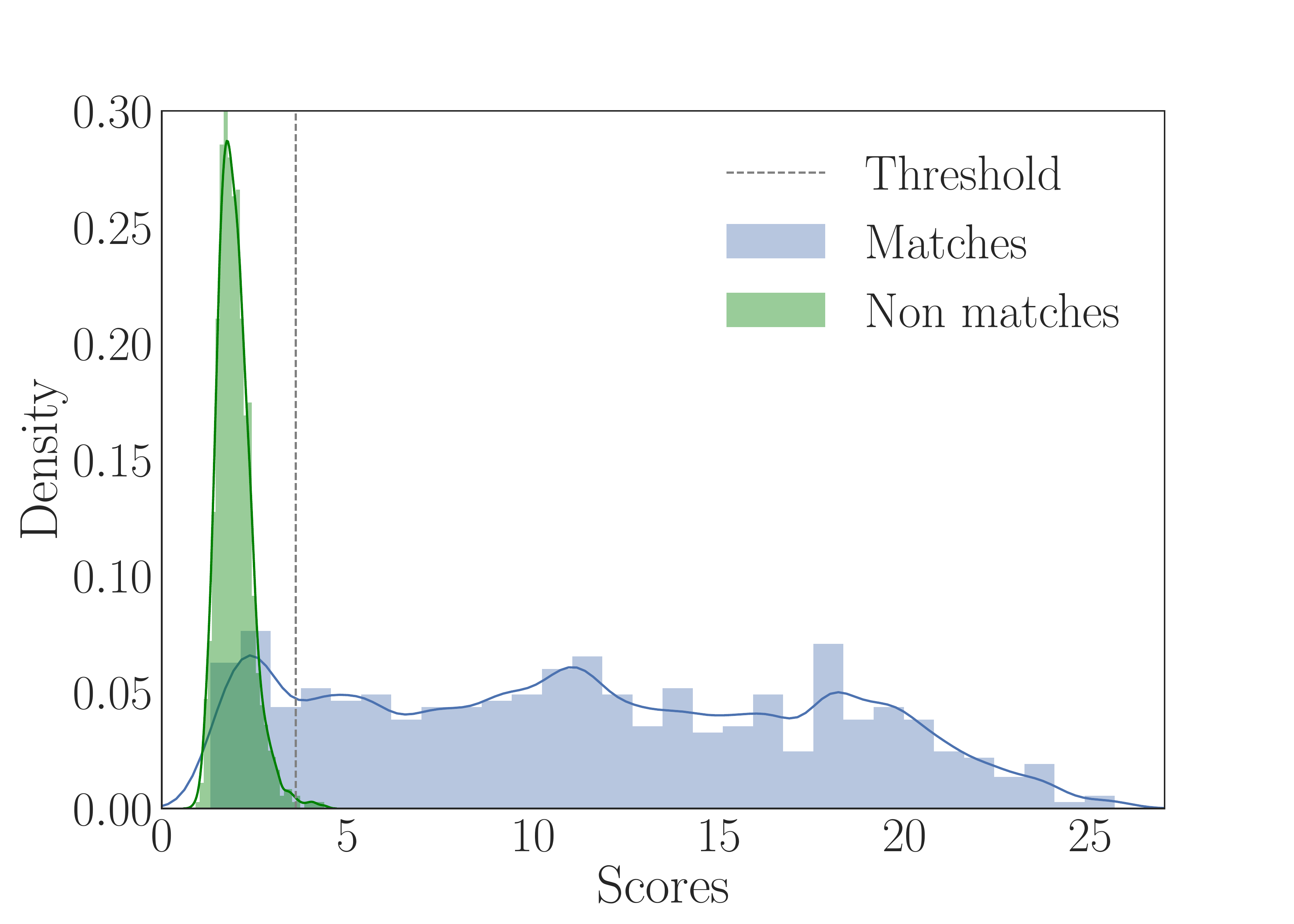}
\caption{\label{fig:histogramscores} Histogram of the scores for positive \textit{(blue)} and negative \textit{(green)} pairs on our internal dataset. The threshold corresponds to the threshold used for Table \ref{table:recall}.}
\end{figure}

Figure \ref{fig:histogramscores} presents the histogram of pairwise scores for all the positive and negative pairs in our internal dataset. 
The distribution of scores for the negative pairs is short-tailed and tightly concentrated around $s=2$. This means that above $s\simeq 5$, all the pairs can be matched with high confidence.  The distribution of scores for the positive pairs is much wider. As we can see from the histogram, a non-negligible fraction of these pairs lies below the classification threshold (dashed vertical line) and thus cannot be detected with this 1-vs-1 method. The purpose of the next section will be to apply an ensemble method to a pool of candidate versions of a given work, to bring these undetected candidates above the threshold by exploiting the many-to-many relationships between the candidates.

\section{Ensemble analysis}
\label{sec:clustering}
While the 1-vs-1 algorithm we presented in Section \ref{sec:1v1} gives satisfying results overall, it still struggles on covers that are significantly different from their original track. Here we show how analyzing a large pool of candidate covers for one given reference track can improve the quality of the matching. The intuition behind this idea is that a cover version can match the reference track poorly, but match another intermediate version which is closer to the reference. For instance, an acoustic cover can be difficult to detect on a 1-vs-1 basis, but might match a karaoke version which itself strongly matches the reference track. We therefore turn to a new use case, where we not only compare single pairs (\textit{e.g.} one reference track against one possible cover), but instead start from a pool of candidates that are all likely to be instances of some given composition (or \emph{work}). Usually, this pool corresponds to candidates that have been pre-filtered according to some non-audio related signal, \textit{e.g.} their title, and might comprise up to a few thousands candidates, depending on the popularity of the work and the specificity of the pre-filtering step. 

\begin{figure}[ht!]
\begin{subfigure}[t]{0.15\textwidth}
    \includegraphics[width=\textwidth]{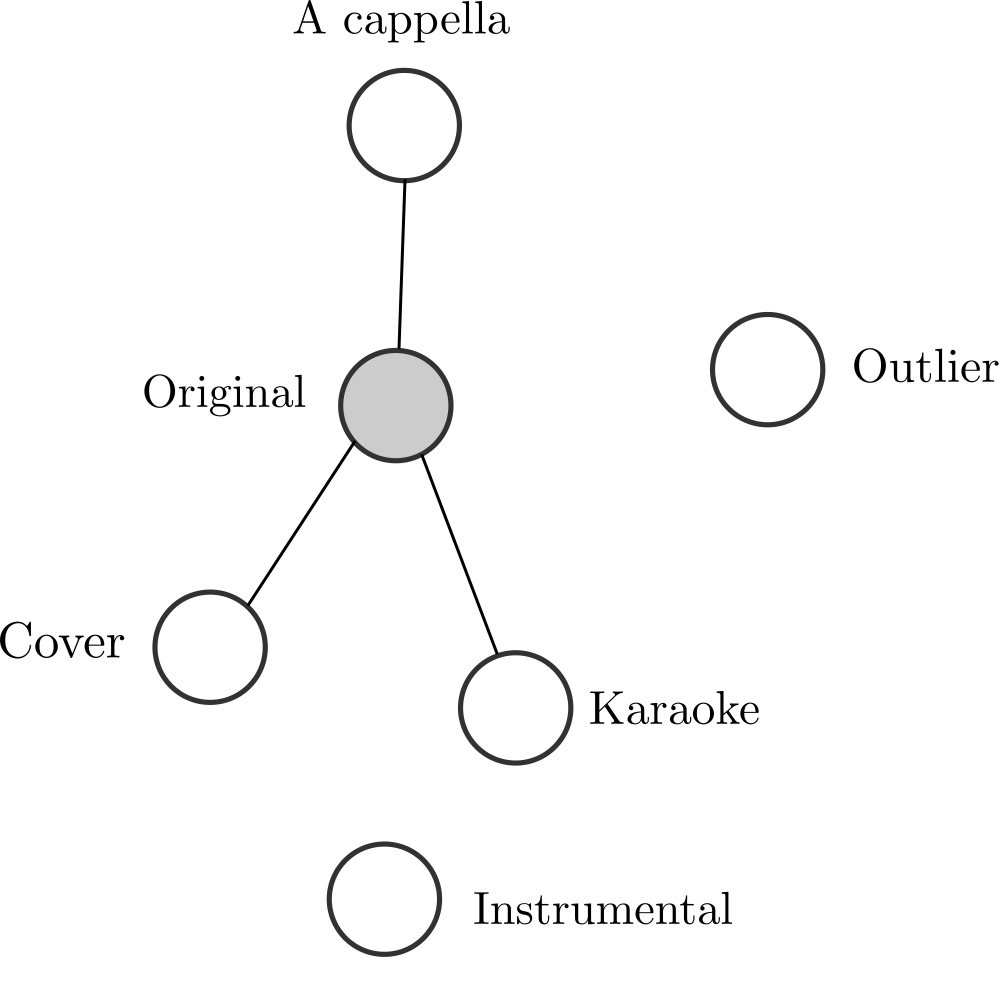}
    \caption{Computing scores versus the reference track}
\end{subfigure}
\rulesep
\begin{subfigure}[t]{0.15\textwidth}
    \includegraphics[width=\textwidth]{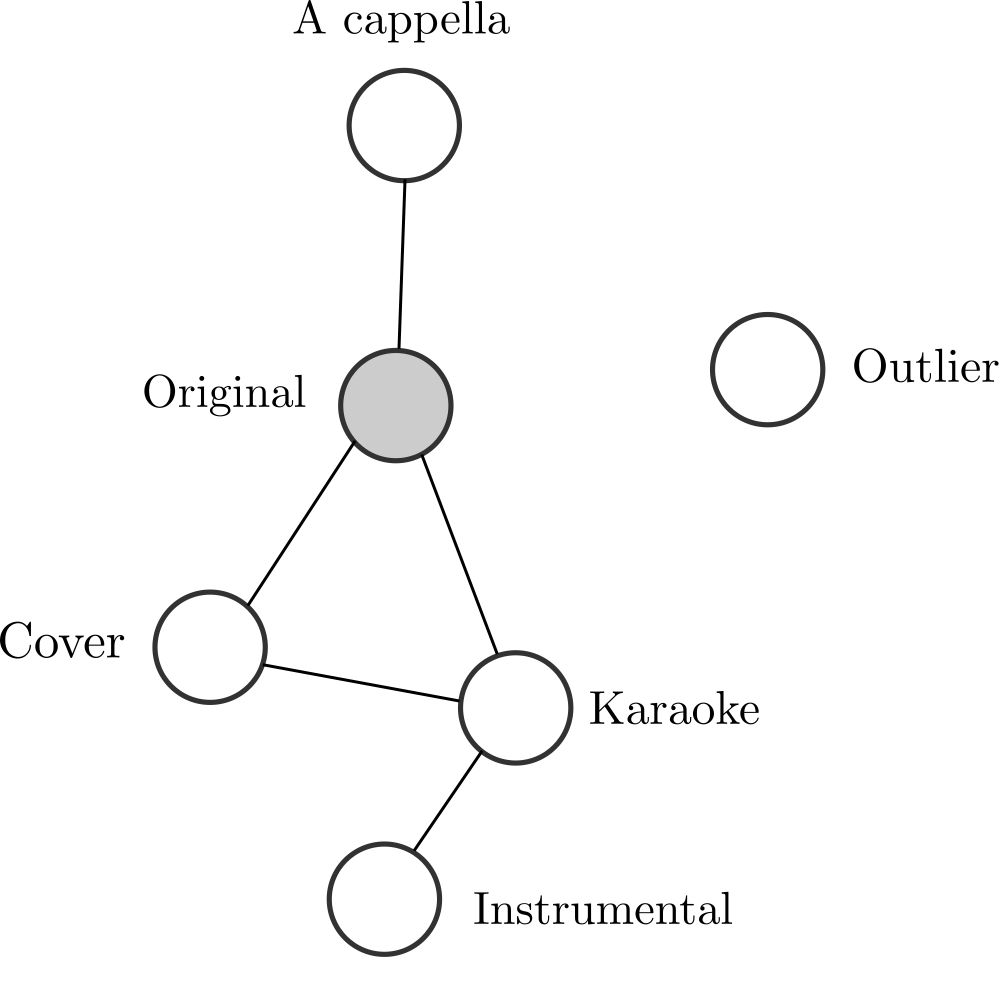}
    \caption{Computing all pairwise scores}
\end{subfigure}
\rulesep
\begin{subfigure}[t]{0.15\textwidth}
    \includegraphics[width=\textwidth]{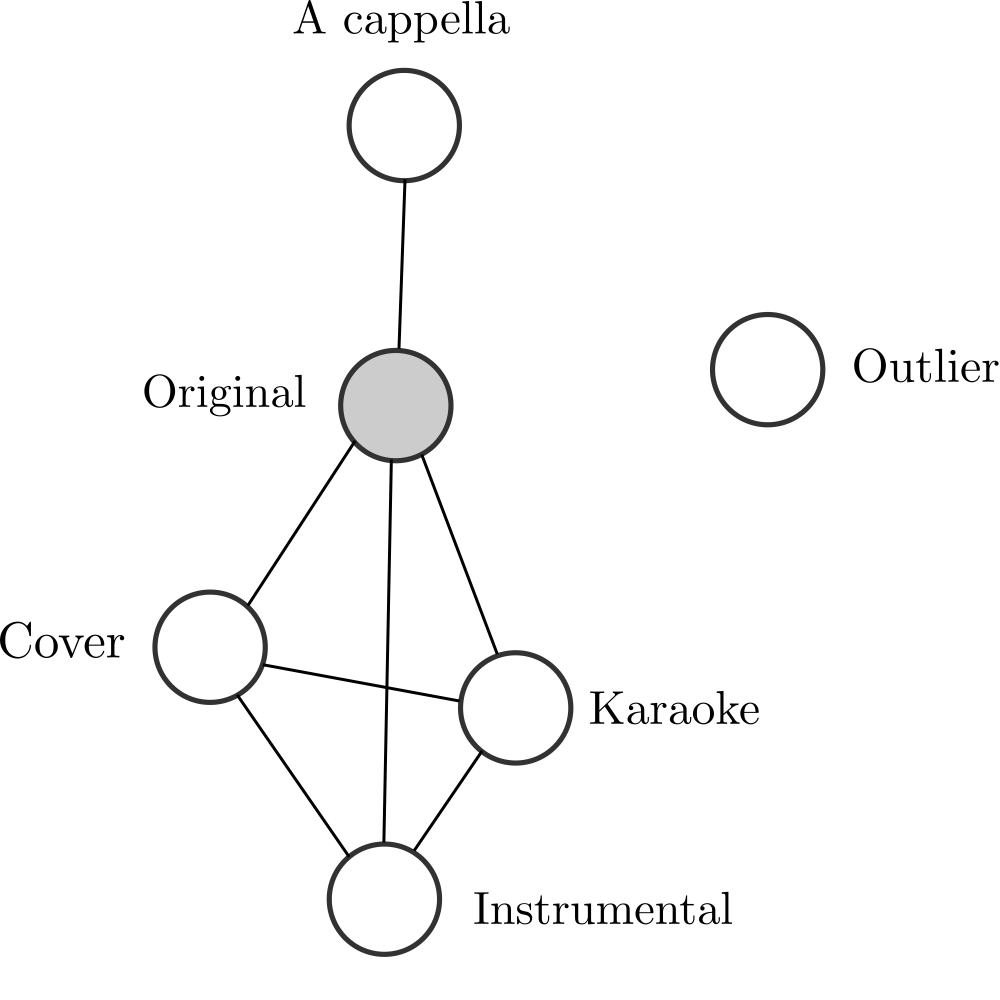}
    \caption{Final graph}
\end{subfigure}
\caption{Direct (a) vs. ensemble-based approach (b)-(c).}
\end{figure}

\subsection{Computing all pairwise scores}\label{sec:NvsN}

Given a set of $N$ candidate versions of a work, we first compare all possible pairs of candidates within the set, resulting in $\frac{N(N-1)}{2}$ distinct scores $\left\{s_{ij}\right\}_{1\leq i < j \leq N}$. As mentioned above, if the candidates have been pre-filtered using some metadata-matching algorithm, $N$ typically varies from a few dozen to a few thousand candidates.



\subsection{Scores to distances}
\label{subsubsec:logistic}


Figure \ref{fig:histogramscores} shows that almost all negative pairs have scores between 0 and 4 while scores above 8 always correspond to positives. Scores above 8 should thus indicate a high probability of a true match regardless of the score, while a variation in score around 4 should have a significant impact on that probability. To account for this fact, we convert our scores into more meaningful distances using a logistic function: $d_{ij} = \left(1 + e^{- \frac{s_{ij}-m}{\sigma}}\right)^{-1}$, where $s_{ij}$ is the  score associated to pair $(i, j)$ and $d_{ij}$ is the resulting distance. We have found that the values $\sigma=0.5$ and $m=4.3$ work well with the distance-collapsing algorithm introduced in the next section. 


\subsection{Collapsing the distances}
\label{subsubsec:collapsing}
\begin{figure}[!ht]
\centering
\begin{minipage}{.21\textwidth}
  \includegraphics[width=\linewidth]{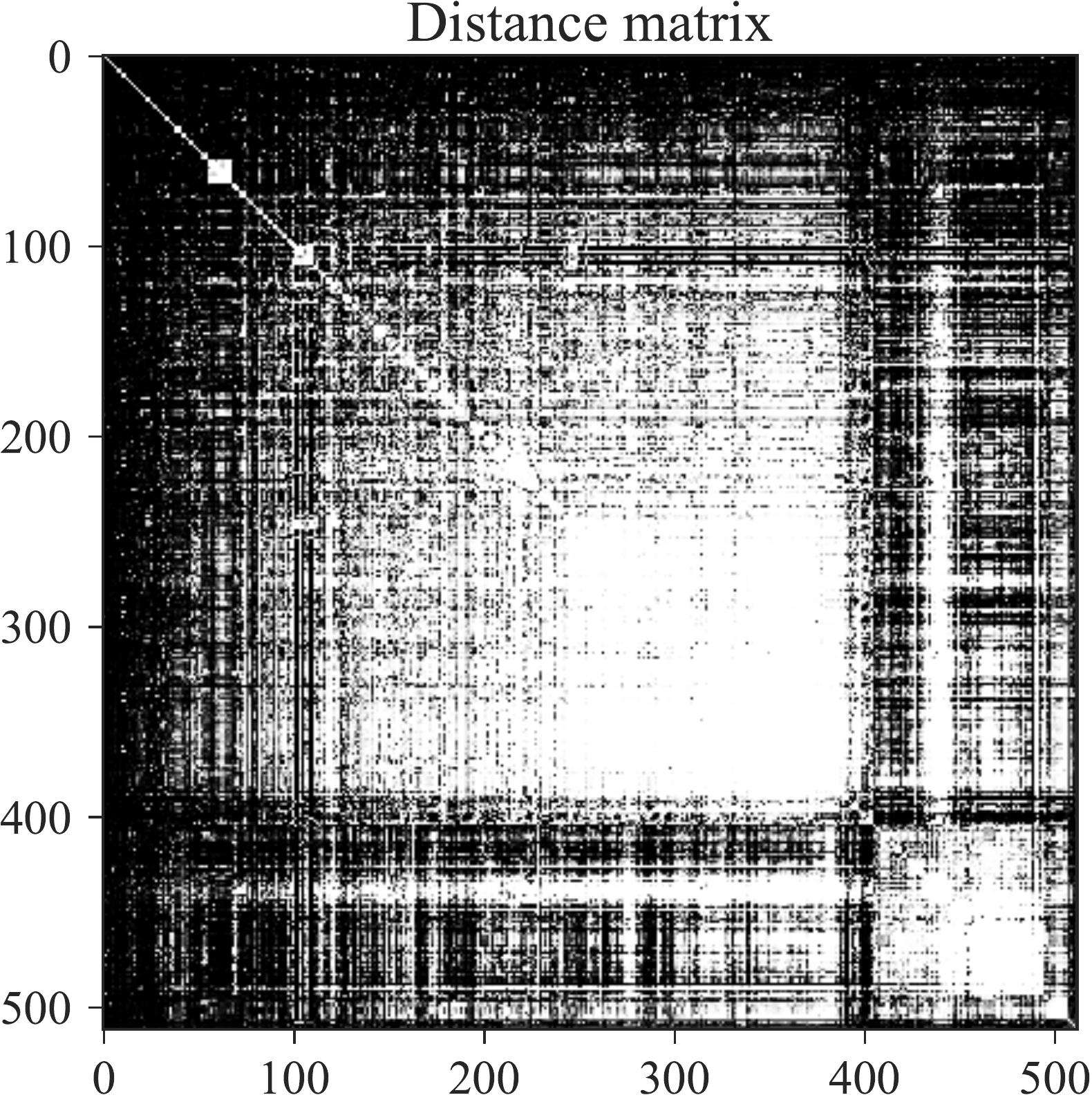}
\end{minipage}%
\begin{minipage}{.21\textwidth}
  \includegraphics[width=\linewidth]{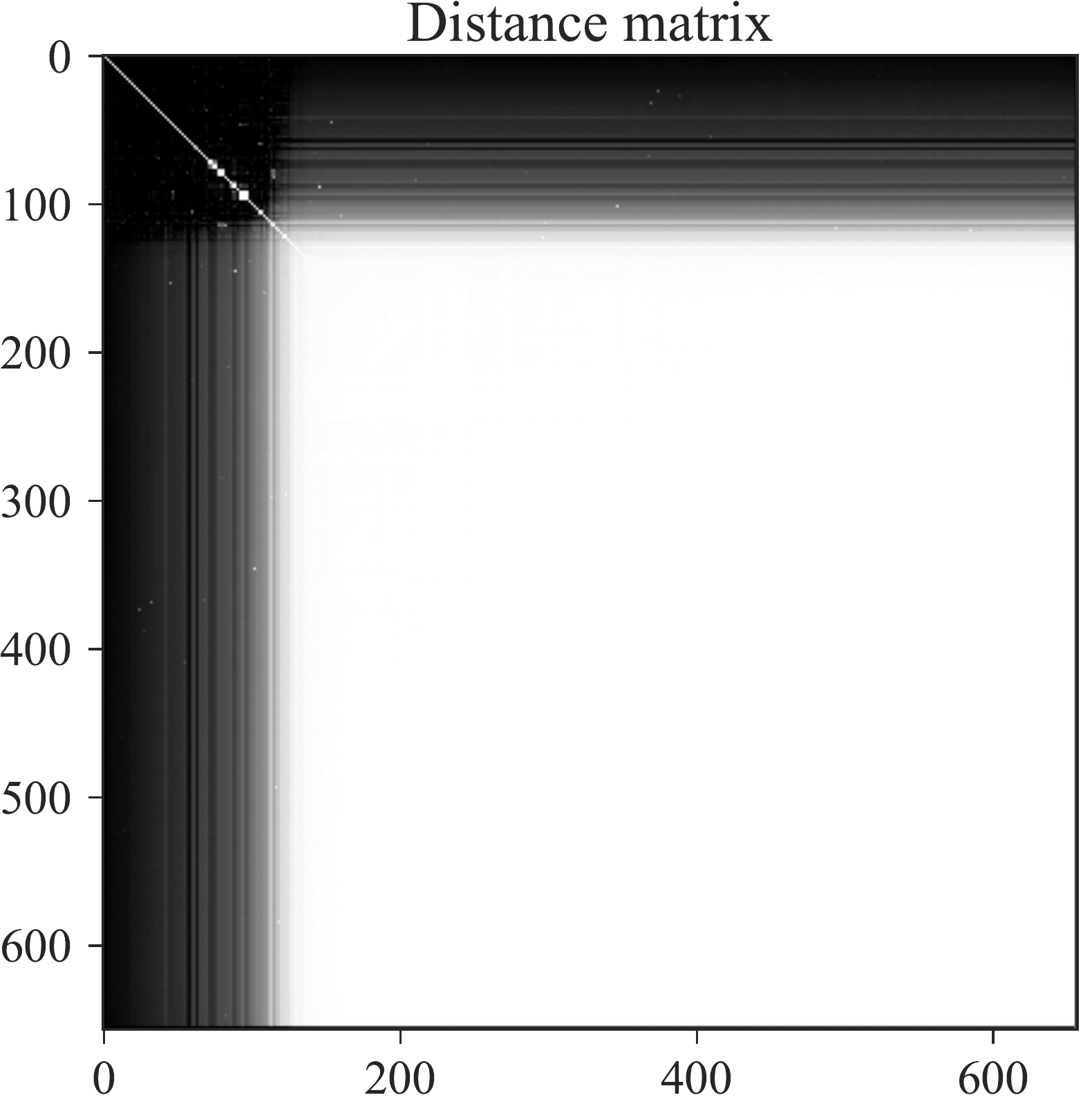}
\end{minipage}
\includegraphics[width=.5\textwidth]{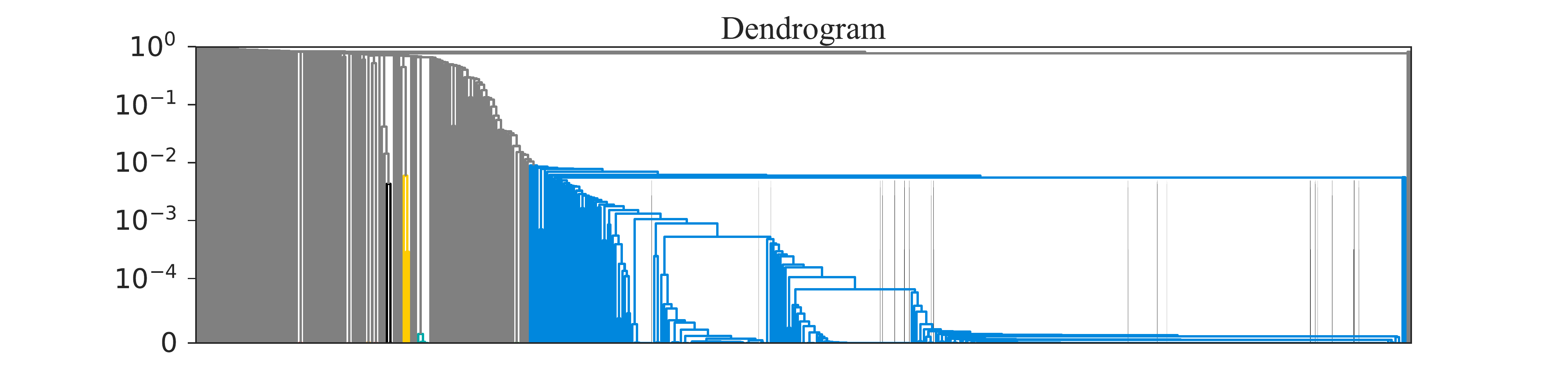}
\caption{\label{fig:floydwarshall} Top: The Floyd-Warshall algorithm applied to the distance matrix of \textit{Get Lucky}, with (left) original distance matrix and (right) the distance matrix after applying the Floyd-Warshall algorithm. Darker shades correspond to larger distances. Bottom: the corresponding dendrogram obtained using hierarchical clustering on the Floyd-Warshall distance matrix.}
\end{figure}

Let $D = \left\{ d_{ij}\right\}$ denote the pairwise distance matrix between all pairs of candidates (see Figure \ref{fig:floydwarshall}, top left). The idea behind the ensemble-based approach is to exploit the geometry of the data to enhance the accuracy of the classification -- for example, the fact that a track can match the reference track better through intermediate tracks than directly. We use a loose version of the Floyd-Warshall algorithm \cite{Floyd:1962:A9S:367766.368168} to update the distances in $D$, such that the new distances satisfy the triangular inequality most of the time\footnote{The distances that would be obtained by applying the original Floyd-Washall algorithm to $D$ would always satisfy the triangular inequality, but the resulting configuration would be very sensitive to outliers. Our method is more robust to outliers, as it requires to find more than one better path to update the distance between two points.}. The method is presented in Algorithm \ref{alg:loosefw}.


\begin{algorithm}
\footnotesize
\caption{\label{alg:loosefw} Loose Floyd-Warshall}
\begin{algorithmic}[1]
\Procedure{CollapseDistances}{distance matrix $D$}
\While {D still updates}
\For{$i, j$ in $1..N$}
\State $\widetilde{D}(i, j) \gets \textbf{min}_{k\neq i, j}^{(2)} D(i, k) + D(k, j) + \eta$
\State $D(i, j) \gets \textbf{min}  \Big(D(i, j),\  \widetilde{D}(i, j) \Big)$
\EndFor
\EndWhile
\EndProcedure
\end{algorithmic}
\end{algorithm}

Here $\mathrm{min}^{(k)}(x)$ denotes the $k^{th}$ smallest value of a vector $x$. We have found that the algorithm is slightly more robust when imposing a penalty $\eta>0$ for using an intermediate node, which we have set to $\eta=0.01$ after performing a grid-search optimization.


Figure \ref{fig:floydwarshall} shows the distance matrix before (top left) and after (top right) updating the distances using Algorithm \ref{alg:loosefw}, for a set of candidates versions of \textit{Get Lucky} by Daft Punk. We can see that the updated distance matrix has a more neatly defined division between clusters of tracks. The figure shows one large cluster in which all tracks are extremely close to each other (the white area), a few smaller clusters (white blocks on the first diagonal) and a number of isolated tracks that match only themselves.


\subsection{Hierarchical clustering}\label{sec:hierarchical}

We then proceed to a clustering of the tracks using the updated distance matrix defined in \ref{subsubsec:collapsing}, denoted $D'$. We use hierarchical clustering as we have no prior knowledge on the number of clusters in the graph. Figure \ref{fig:floydwarshall} (bottom) shows the dendrogram associated with the hierarchical clustering applied to $D'$. In this example, if we apply a relatively selective threshold, we find one major cluster (colored in blue in Figure \ref{fig:floydwarshall}) that contains 97\% of the true positives and no false positives. Most other clusters contain a single element, which are all the negative tracks and the remaining 3\% of the positives.
If we set the clustering threshold lower, then we can get more granular clusters within a same work. 

\subsection{Final score}\label{sec:cophenetic}

In order to assign each track a final score that measures its similarity to the reference track, we use the \textit{cophenetic distance} to the reference track, \textit{i.e.} the distance along the dendrogram that is produced by the hierarchical clustering. Each track is thus assigned a final score in $0-100$, simply taken equal to 100 $\times$ (1 - cophenetic distance), such that exact matches have a score of 100.

\section{Analysis of real world examples}
\label{sec:results}

\subsection{Data}
We now apply the above to real world data. Our dataset consists of 10 sets of candidates that correspond to 10 works that we want to find the versions of. These 10 works span multiple genres and musical styles, including Hip Hop, R\&B, Rap, Pop and Jazz. For a given work, we create the set of candidates by performing a metadata search of the given work's title on the whole Spotify catalogue. Across the given works that we study, this produces sets of candidates whose sizes vary from a few hundred to a few thousand candidate tracks. 
Each set includes a reference track, which will be the anchor point for that composition. More details on the dataset can be found in Table \ref{dataset}.

\begin{table}[h]
\small
\begin{center}
\setlength{\tabcolsep}{0.8ex}
\begin{tabular}{| c | c | c | c | c |}
  \hline			
  \textbf{Work} & \textbf{\# tracks} & \textbf{\% positives} & \begin{tabular}[x]{@{}c@{}}\textbf{Reference}\\\textbf{artist}\end{tabular} \\ 
  \hline  
  \href{https://open.spotify.com/track/6lV2MSQmRIkycDScNtrBXO?si=xdSaKceZQS-MqGR8O1J4Og}{\textit{Airplane}}
  & 811 & 19\% & B.o.B 
  \\
  \href{https://open.spotify.com/track/05KfyCEE6otdlT1pp2VIjP?si=nZRyhDndSH6rBmnaRMiMnA}{\textit{Believer}}
  & 2552 & 5\%  & Imagine Dragons 
  \\
   \href{https://open.spotify.com/track/5PUvinSo4MNqW7vmomGRS7?si=HoWWVLjFSd6V2MpySx-JdA}{\textit{Blurred Lines}}
   & 386 & 71\% & Robin Thicke 
  \\
  \href{https://open.spotify.com/track/2771LMNxwf62FTAdpJMQfM?si=QjthOI5dSTy5FZ_WMUXQSg}{\textit{Bodak Yellow}}
  & 110 & 78\% & Cardi B
  \\
  \href{https://open.spotify.com/track/7rt0kEDWRg3pgTZJKuszoE?si=zGZGkIM0RSmp5FYtHIbBaQ}{\textit{Brown Sugar}}
  & 721 & 5.8\% & D'Angelo 
  \\
  \href{https://open.spotify.com/track/00gLKa8SbzW3SlRwlfh6U6?si=L6AH33NDROeOnp8wNDN3rQ}{\textit{Embraceable You}}
  & 1319 & 94\% & Sarah Vaughan 
  \\
  \href{https://open.spotify.com/track/2Foc5Q5nqNiosCNqttzHof?si=qCa_9l5mTJGxmF-lR_No3A}{\textit{Get Lucky}}
  & 657 & 83\% & Daft Punk  
  \\
  \href{https://open.spotify.com/track/0DCdMiGUOlZM8YRRF2kzR7?si=mxnfznyrQEuukrOVQmCA9w}{\textit{Halo}}
  & 2995 & 7.9\% & Beyoncé  
  \\ 
  \href{https://open.spotify.com/track/28siypca4TEqLnQ6Cgbdbe?si=YJO0kJF2QQyA4wallfNwug}{\textit{Heartless}} 
  & 1747 & 5.3\%  & Kayne West  
  \\
  \href{https://open.spotify.com/track/7pKfPomDEeI4TPT6EOYjn9?si=OsgDBo7yRwu3NtL_6FE47Q}{\textit{Imagine}}
  & 2044 & 50\% & John Lennon 
  \\
   \hline  
  \end{tabular}
 \end{center}
 \caption{\label{dataset} The ``10 works'' dataset. For each work, we have selected a reference track that will be our anchor point for that composition. 
 \textit{Click on a work to play the reference track in the browser}.}
\end{table}


\subsection{Outline of the analysis}

For each of these works, we analyze the set of candidates following the steps outlined in the previous two sections, providing us with two sets of outputs for each work: \textit{\textbf{(a)}} the \textit{direct score}, defined as the output of the 1-vs-1 algorithm between each candidate and the reference track, as described in Section \ref{sec:1v1} (rescaled between 0 and 100); \textit{\textbf{(b)}} the \textit{ensemble-based score}, produced by the method described in Section \ref{sec:clustering} (also between 0 and 100).

In the next section we start by quantitatively evaluating our ensemble-based approach (b) against the direct approach (a), before turning to some qualitative examples.

\subsection{Quantitative results}\label{sec:metrics}

We define two different metrics to evaluate the direct and the ensemble-based methods:


\textbf{Ranking metric:} For each work, we pick the value of the threshold that minimizes the number of classification errors, and report the number of errors. We call this a \textit{ranking metric} as the number of errors is minimized when positives and negatives are perfectly ranked, regardless of their scores. 
We also report the corresponding recall and false positive rates for this threshold.

\textbf{Classification metric:} We fix a universal classification threshold and compute the corresponding number of classification errors. 

\begin{table}[h!]
\footnotesize
\begin{center}
\setlength{\tabcolsep}{0.3ex}
\begin{tabular}{| c | c | c | c | c | c | c | c |}
  \hline			
   &&\multicolumn{6}{c |}{\textbf{Ranking errors - direct}}\\
  \cline{3-8}
  \textbf{Work} & \textbf{Best thr.} & \multicolumn{2}{c |}{False negatives} & \multicolumn{2}{c |}{False positives} & \multicolumn{2}{c |}{Both} \\
  \cline{3-8}
  && Abs. & Rel. & Abs. & Rel. & Abs. & Rel. \\
  \hline
  \textit{Airplane} &  12.1 & 33 & 21.9\% & 1 & 0.2\% & 34 & 4.2\%  \\
  \textit{Believer} &  18.2 & 6 & 5.2\% & 0 & 0.0\% & 6 & 0.2\%  \\
  \textit{Blurred Lines} & 10.1 & 19 & 7.0\% & 9 & 8.3\% & 28 & 7.3\%  \\
  \textit{Bodak Yellow} &  6.1 & 6 & 7.0 \%& 6 & 33.3\% & 12 & 10.9\%  \\
  \textit{Brown Sugar} &  12.1 & 2 & 4.8\% & 1 & 0.1\%  & 3 & 0.4\%  \\
  \textit{Embraceable You} &  4 & 0 & 0\% & 74 & 98.7 \%  & 74  & 5.6 \% \\
  \textit{Get Lucky} &10.1 & 17 & 3.1\% & 3  & 2.6\% & 20 & 3.0\%  \\
  \textit{Halo} & 11.1 & 8 & 3.4\% & 8 & 0.3\%  & 16 & 0.5\% \\
  \textit{Heartless} & 12.2 & 15 & 16.3\% & 2 & 0.1\%  & 17 & 1.0\%  \\
  \textit{Imagine} &  15.2 & 72 & 7.1\% & 17 & 1.7\% & 89 & 4.4\%  \\
   \hline  
  \end{tabular}
	\subcaption{\label{table:results-ranking-direct} Direct approach.}

  \vspace{10pt}
  
\begin{tabular}{| c | c | c | c | c | c | c | c |}
  \hline			
   &&\multicolumn{6}{c |}{\textbf{Ranking errors - ensemble-based}}\\
  \cline{3-8}
  \textbf{Work} & \textbf{Best thr.} & \multicolumn{2}{c |}{False negatives} & \multicolumn{2}{c |}{False positives} & \multicolumn{2}{c |}{Both} \\
  \cline{3-8}
  && Abs. & Rel. & Abs. & Rel. & Abs. & Rel. \\
  \hline
  \textit{Airplane} & 70.7 & 4 & 2.6\% & 3 & 0.5\% & 7 & 0.9\%  \\
  \textit{Believer} & 85.9 & 0 & 0.0\% & 0 & 0.0\% & 0 & 0.0\%  \\
  \textit{Blurred Lines} & 52.5 & 0 & 0.0\% & 0 & 0.0\% & 0 & 0.0\%  \\
  \textit{Bodak Yellow} & 29.3 & 9 & 10.5\% & 0 & 0.0\% & 9 & 8.2\%  \\
  \textit{Brown Sugar} & 70.7 & 0 & 0.0\% & 1  & 0.1\% & 1 & 0.1\%  \\
  \textit{Embraceable You} & 40.4 & 22 & 1.8 \% & 19 & 25.3 \% & 41 & 3.1 \% \\
  \textit{Get Lucky} & 78.8 & 5 & 0.9\%  & 2 & 1.7\% & 7 & 1.1\%  \\
  \textit{Halo} & 98.0 & 4 & 1.7 \%& 20  & 0.7\% & 24 &  0.8\% \\
  \textit{Heartless} &  83.8 & 8 & 8.7\% & 1  & 0.1\% & 9 & 0.5\%  \\
  \textit{Imagine} & 96.0 & 1 & 0.1\% & 5 & 0.5\% & 6 & 0.3\%  \\
   \hline  
  \end{tabular}
 	\subcaption{\label{table:results-ranking-graphical} Ensemble-based approach.}

 \end{center}
  \caption{\label{results:A} Optimal thresholds and corresponding results for the ranking metric. 
  }
\end{table}

Table \ref{results:A} shows the results for the ranking metric for each work in our dataset. For the optimal thresholds, we report the number of false negatives, false positives and the sum of both (\textit{i.e.} the total number of classification errors). We also compute the corresponding false negative rate, false positive rate and total error rate. 

Table \ref{table:results-ranking-direct} shows the ranking results for the direct approach. Interestingly, the number of false negatives tends to be higher than the number of false positives.\footnote{\textit{Embraceable You} is an exception, as its threshold is degenerate and all tracks are classified as matching.} This is in line with the histogram in Figure \ref{fig:histogramscores}, which shows a short-tailed distribution for the negatives and a wider distribution for the positives. Overall, the error rate lies between $0-10\%$, corresponding to a recall rate between $80\%$ and $97\%$ and a false positive rate below $10\%$ (except for \textit{Bodak Yellow} and \textit{Embraceable You} which have a very small number of negatives to begin with -- the latter case is in fact degenerate as nearly all tracks are classified as matching). 

Table \ref{table:results-ranking-graphical} shows the ranking results for our ensemble-based approach. The number of ranking errors is substantially lower than for the direct approach, including both the number of false positives and false negatives, as the total error rate goes down below $1\%$ in most cases. Again, the main exception is \textit{Bodak Yellow}, which has the smallest number of candidates.\footnote{It was also a genuinely difficult example and we struggled to annotate it.} \textit{Embraceable You} is the second most challenging work, but remarkably its threshold is no longer degenerate, meaning that the method has now found a way to separate the candidates. Notably, the number of false negatives no longer outnumbers the number of false positives: the ensemble-based approach has successfully caught most of the difficult tracks that poorly matched the reference track. Among the few tracks that are still missed, several are actually very close to the threshold, and only a handful are still completely undetected (cf Table \ref{misses}).  



\begin{table}[h!]
\footnotesize
\begin{center}
\setlength{\tabcolsep}{0.6ex}
\begin{tabular}{| c | c | c | c | c | c | c | c |}
  \hline			
   &&\multicolumn{6}{c |}{\textbf{Classification errors - direct}}\\
  \cline{3-8}
  \textbf{Work} & \textbf{Thr.} & \multicolumn{2}{c |}{False negatives} & \multicolumn{2}{c |}{False positives} & \multicolumn{2}{c |}{Both} \\
  \cline{3-8}
  & & Abs. & Rel. & Abs. & Rel. & Abs. & Rel. \\
  \hline
  \textit{Airplane} & 12.1 & 33 & 21.9\% & 1 & 0.2\% & 34 & 4.2\%  \\
  \textit{Believer} & 12.1 & 4 & 3.5\% & 91 & 3.7\% & 95 & 3.7\%  \\
  \textit{Blurred Lines} & 12.1 & 28 & 10.3\% & 0 & 0\% & 28 & 7.3\%  \\
  \textit{Bodak Yellow} & 12.1 & 49 & 57 \%& 0 & 0\% & 49 & 44.5\%  \\
  \textit{Brown Sugar} & 12.1 & 2 & 4.8\% & 1 & 0.1\%  & 3 & 0.4\%  \\
  \textit{Embraceable You} & 12.1 & 753 & 60.5 \% & 3 & 4 \% & 756 & 57.3 \%  \\
  \textit{Get Lucky} & 12.1 & 23 & 4.2\% & 1 & 0.9\% & 24 & 3.7\%  \\
  \textit{Halo} & 12.1 & 12 & 5.1\% & 4 & 0.1\%  & 16 & 0.5\% \\
  \textit{Heartless} & 12.1 & 15 & 16.3\% & 2 & 0.1\%  & 17 & 1.0\%  \\
  \textit{Imagine} & 12.1 & 49 & 4.8\% & 94 & 9.3\% & 143 & 7\%  \\
   \hline  
  \end{tabular}
	\subcaption{\label{table:results-classification-direct} Direct approach.}
    
  \vspace{10pt}
\begin{tabular}{| c | c | c | c | c | c | c | c |}
  \hline			
   &&\multicolumn{6}{c |}{\textbf{Classification errors - ensemble-based}}\\
  \cline{3-8}
  \textbf{Work} & \textbf{Thr.} & \multicolumn{2}{c |}{False negatives} & \multicolumn{2}{c |}{False positives} & \multicolumn{2}{c |}{Both} \\
  \cline{3-8}
  & & Abs. & Rel. & Abs. & Rel. & Abs. & Rel. \\
  \hline
  \textit{Airplane} & 78.8 & 9 & 6.0\% & 1 & 0.2\% & 10 & 1.2\%  \\
  \textit{Believer} & 78.8 & 0 & 0.0\% & 27 & 1.1\% & 27 & 1.1\%  \\
  \textit{Blurred Lines} & 78.8 & 2 & 0.7\% & 0 & 0.0\% & 2 & 0.5\%  \\
  \textit{Bodak Yellow} & 78.8 & 19 & 22.1\% & 0 & 0.0\% & 19 & 17.3\%  \\
  \textit{Brown Sugar} & 78.8 & 2 & 4.8\% & 1  & 0.1\% & 3 & 0.4\%  \\
  \textit{Embraceable You} & 78.8 & 70 & 5.6 \% & 1 & 1.3 \% & 71 & 5.4 \% \\
  \textit{Get Lucky} & 78.8 & 5 & 0.9\%  & 2 & 1.7\% & 7 & 1.1\%  \\
  \textit{Halo} & 78.8 & 0 & 0 \%& 163  & 5.9\% & 163 &  5.4\% \\
  \textit{Heartless} &  78.8 & 7 & 7.6\% & 4  & 0.2\% & 11 & 0.6\%  \\
  \textit{Imagine} & 78.8 & 0 & 0\% & 158 & 15.6\% & 158 & 7.7\%  \\
   \hline  
  \end{tabular}
  	\subcaption{\label{table:results-classification-graphical} Ensemble-based approach.}

 \end{center}
 \caption{\label{results:B} Universal threshold and corresponding results for the classification metric. 
 }
\end{table}

Table \ref{results:B} shows the results for the classification metric. The universal threshold for each approach is defined as the median of the optimal thresholds obtained in the ranking experiment above. Again, we report the number of false negatives, the number of false positives and the sum of both. We also compute the corresponding false negative rate, false positive rate and total error rate. Here again, the results of the ensemble-based approach are overall superior to the direct approach, mostly due to an increase in recall. 
Although Table \ref{table:results-classification-direct} is quite similar to Table \ref{table:results-ranking-direct}, which is a sign that the threshold on direct scores can be chosen in a nearly universal way, Table \ref{table:results-classification-graphical} differs considerably from Table \ref{table:results-ranking-graphical} for some specific works (namely \textit{Halo}, \textit{Imagine} and \textit{Believer}). This happens as the optimal threshold is significantly higher on these works (often $>95\%$), letting a large number of false positives above the 78.8\% threshold. 


\subsection{Examples}\label{sec:examples}

For each work, we can identify the cases where the ensemble-based approach has allowed us to detect previously undetected tracks, and trace back the optimal path that joined the reference track and the newly found track. Table \ref{paths} shows a few examples of such paths for various works. For each example, the reference track is shown at the top of the cell (depth 0), and the newly found track at the bottom of the cell (depth $>1$), with the intermediate tracks that allowed to bridge the gap in between. All the examples are true positives, except for the last example (\textit{Halo Halo} by Fajters), which has been erroneously matched to a karaoke version of the reference track.


\begin{table}[t!]
\footnotesize
\begin{center}
\setlength{\tabcolsep}{0.56ex}
\begin{tabular}{| c | c | c | c | c | c | c |}
  \hline			
\textbf{Work} & \textbf{Depth} & \textbf{Main artist (\& link)} & \begin{tabular}[x]{@{}c@{}}\textbf{Direct}\\\textbf{scores} \end{tabular} & \begin{tabular}[x]{@{}c@{}}\textbf{Ensemble-based}\\\textbf{scores}\end{tabular}  \\
  \hline	
					& 0 & \href{https://open.spotify.com/track/7pKfPomDEeI4TPT6EOYjn9?si=OsgDBo7yRwu3NtL_6FE47Q}{John Lennon} & \textbf{100} & \textbf{100} \\
\textit{Imagine}    & 1 & \href{https://open.spotify.com/track/4szK84IjHD9sYi2acYQuy2?si=HVjmfKf3RqW6AfE1BHvsgg}{Classic Gold Hits} & \textbf{60.0} & \textbf{99.99} \\
					& 2 & \href{https://open.spotify.com/track/066N0phJeNFLEiREiP65VG?si=EaCo5Pa1Ss6bjAyz4M2-NQ}{A Perfect Circle} & \textbf{21.6} & \textbf{97.9} \\
					& 3 & \href{https://open.spotify.com/track/4CZdBLanxIhSElTxSpIH5b?si=JUXL1Rk0R2iKhcT4r-_VyA}{Yoga Pop Ups} & 8.6 & \textbf{97.9} \\
   \hline
 					& 0 & \href{https://open.spotify.com/track/28siypca4TEqLnQ6Cgbdbe?si=YJO0kJF2QQyA4wallfNwug}{Kanye West} & \textbf{100} & \textbf{100} \\
\textit{Heartless}  & 1 & \href{https://open.spotify.com/track/0DU2iZx61j4QCsjEJQHeS8?si=zR_4bZH0RHSSm_dzsIFkuA}{The Fray} & \textbf{34.1} & \textbf{99.85} \\
 	   		        & 2 & \href{https://open.spotify.com/track/1OV00OgzTFS8JkAmd7ABR6?si=BKJHJ9XjSiORb8Htto-PyA}{William Fitzsimmons} & 9.5 & \textbf{90.7} \\
   \hline	         
 					& 0 & \href{https://open.spotify.com/track/2Foc5Q5nqNiosCNqttzHof?si=qCa_9l5mTJGxmF-lR_No3A}{Daft Punk} & \textbf{100} & \textbf{100} \\
\textit{Get Lucky}  & 1 & \href{https://open.spotify.com/track/6fclktfmcORXSsD4Acl5HF?si=p8BhHTc8RLSiG8qwhKgfvQ}{Samantha Sax} & \textbf{40.4} & \textbf{99.95} \\
 	         		& 2 & \href{https://open.spotify.com/track/0JvSslOE8D16WpkCkABzNE?si=c7W3UB9tTPe8yXvZj2kAwA}{Dallas String Quartet} & 6.9 & \textbf{86.6} \\
   \hline	        
 					& 0 & \href{https://open.spotify.com/track/0DCdMiGUOlZM8YRRF2kzR7?si=mxnfznyrQEuukrOVQmCA9w}{Beyonce} & \textbf{100} & \textbf{100} \\
\textit{Halo}		& 1 & \href{https://open.spotify.com/track/35LlWyY6u5ajhNbHkalXkB?si=PBR2fzq6TfabzKtZGzBIgw}{LP}  & \textbf{27.6} & \textbf{99.96} \\
 	         		& 2 & \href{https://open.spotify.com/track/3e9WMvJgfFzHy7KDpqJQqh?si=wOM5OYiTTiW0BSHvFigy8A}{Dion Lee} & 7.96 & \textbf{99.16} \\
   \hline
\textit{Halo}		& 0 & \href{https://open.spotify.com/track/0DCdMiGUOlZM8YRRF2kzR7?si=mxnfznyrQEuukrOVQmCA9w}{Beyonce} & \textbf{100} & \textbf{100} \\
  					& 1 & \href{https://open.spotify.com/track/0ditbVz3cyzR6BQrxcKvOA?si=nQC6vL9aSU2wZTcW5NcCKw}{Karaoke Universe}  & \textbf{20.5} & \textbf{99.96} \\
\textit{Halo Halo}	& 2 & \href{https://open.spotify.com/track/2EO4dhWzJG8HCL4hNgTQW4?si=JgN16LRnSlyIyseCxASX2Q}{Fajters} & 7.45 & \textbf{99.35} \\
   \hline	
     \end{tabular}
   \end{center}
   \caption{\label{paths} Some examples of tracks that are undetected by the direct approach and captured by the ensemble-based approach, in the ranking experiment. The scores that are above the detection thresholds for each method are displayed in bold (the corresponding detection thresholds can be found in Table \ref{results:A}). \textit{Click on an artist to play in the browser}.}
  \end{table}

What about the tracks that are still undetected? Table \ref{misses} shows examples of tracks that are still undetected by our ensemble-based approach for a couple of works. No clear pattern  emerges -- apart from the fact that they are often in a very different musical style from the original.

\begin{table}[t!]
\footnotesize
\centering
\setlength{\tabcolsep}{0.8ex}
\begin{tabular}{| c | c | c | c | c | c |}
  \hline
\textbf{Work}& \textbf{Main artist - Title} & \begin{tabular}[x]{@{}c@{}}\textbf{Direct}\\\textbf{score} \end{tabular} & \begin{tabular}[x]{@{}c@{}}\textbf{Ensemble}\\-\textbf{based}\\\textbf{score} \end{tabular}  \\
  \hline	
  \textit{Get Lucky} & The Getup - \href{https://open.spotify.com/track/3W944t5PM0e7cuSj9k01XE?si=Rh_KlK1KT76qQGymsVuvaA}{\textit{Get Lucky}} & 6.4 & 24.9  \\
   \hline	
  		 \textit{Halo} & Polina Kermesh - \href{https://open.spotify.com/track/562PPTMnlQkDt3O4rxJFmA?si=nxaC9mg7RCW1s-GrEX7txA}{\textit{Halo}} & 6.3 & 98.1 \\
  		& Amanda Sense - \href{https://open.spotify.com/track/356d9i2vsYS7zDCg2HATtg?si=uhstl0X9R6uVKO1s4DDX1w}{\textit{Halo}} & 12.4 & 94.3 \\
   \hline
  \textit{Imagine} & Dena De Rose - \href{https://open.spotify.com/track/51x7Ia264SWiSmLq1YB1sN?si=RoiFmfG3SHKVVWJBKvoXVg}{\textit{Imagine}} & 10.4 & 86.2 \\
   \hline
 \textit{Embraceable}	& Earl Hines  - \href{https://open.spotify.com/track/1Kos7DKwmn2frHqZtS72b6}{\textit{Embraceable You}} & 9.6 & 36.3 \\
  \textit{You} & Samina  - \href{https://open.spotify.com/track/29INEYnk37kJ9PhY4IZYLr}{\textit{Embraceable You}} & 5.8 & 17.0 \\
   \hline
   	\textit{Heartless} & Bright Light - \href{https://open.spotify.com/track/5KFmbjXswQefMej9noa35t?si=AyBChL4HTICaosAd4R39Ow}{\textit{Heartless}} & 11.9 & 50.3 \\
   & Rains - \href{https://open.spotify.com/track/01WuRGzdHatPEn2oxiYaQG?si=DmQxSGafQtyyMxsTx5shZA}{\textit{Heartless}} & 6.4 & 47.7 \\
   \hline	
  \textit{Bodak Yellow}	& Josh Vietti - \href{https://open.spotify.com/track/2MLP51NFC6RKSvDRWj8QEX?si=TF1ifhTZRxSYYjUNKSsVkw}{\textit{Bodak Yellow}} & 5.8 & 14.2 \\
         & J-Que Beenz - \href{https://open.spotify.com/track/4NP1uXIqS5dEbxpEIELNP8?si=ZKbJx3DkS-ePUyYLiwPf_A}{\textit{Bodak Yellow}} & 5.6 & 13.0	\\
   \hline  
   \textit{Airplanes} & Em Fresh - \href{https://open.spotify.com/track/3zdHo8xxdM7ovkr2gBQAE6?si=a7z0epr-QFWgeAv54x_bSg}{\textit{Airplanes}} & 5.5 & 66.1 \\
         & Lisa Scinta - \href{https://open.spotify.com/track/50tsm5izJLx7Of5z8l96yG?si=ZFaLKmteQuivw2fUFFhKWA}{\textit{Airplanes}} & 9.6 & 55.0 \\
     \hline  
  \end{tabular}
   \caption{\label{misses} Some example of tracks that are undetected by the ensemble-based approach in the ranking experiment, with their scores for both methods. \textit{Click on a title to play in the browser}.}
  \end{table}


\section{Discussion}\label{sec:discussion}

One main challenge associated with our ensemble-based approach is how to correctly handle transitivity. This issue emerges from the fact that compositions are not mutually exclusive. For example, a medley might constitute a bridge between two distinct composition groups, which our algorithm would then merge together (which is undesirable). There are probably at least two ways around this issue: one is metadata-based (\textit{i.e.} identify these potential outliers from the metadata and exclude them from the graph computation), while another is to detect them directly from the graph structure (identify bridges between otherwise unrelated clusters).

\section{Conclusion}\label{sec:conclusion}

In this paper, we have introduced a new formulation of the cover song identification problem: among a pool of candidates that are likely to match one given reference track, find the actual positives. We have introduced a two-step approach, with a first step that computes pairwise similarities between every pair of tracks in the pool of candidates (for which any known 1-vs-1 approach can be used), and a second ensemble-based step that exploits the relationships between all the candidates to output final results. We have shown that this second step can significantly improve the performance compared to a pure 1-vs-1 approach, in particular on the ranking task, where the error rate is down from a few percents to less than 1\% in general.  The classification task is naturally more challenging as the optimal threshold might vary from work to work, suggesting that the method would be best exploited as a complement to human annotations -- where the human's task would mainly be to find the optimal threshold for the classification. Automating this last step turned out to be non-trivial and is left for future work.

\bibliography{track-to-work-ismir}

%
%
%
%

\end{document}